\pdfoutput=1

\documentclass[fleqn,10pt]{wlpeerj}

\usepackage{graphicx}
\usepackage{booktabs}
\usepackage{hyperref}
\usepackage{amsmath,amssymb}
\usepackage{xcolor}

\hypersetup{
    colorlinks=true,
    linkcolor=blue,
    urlcolor=blue,
    citecolor=blue
}

\title{Pre-AI Baseline: Developer IDE Satisfaction and Tool Autonomy in 2022}

\author[1]{Nikola Balić}
\affil[1]{Faculty of Science, University of Split, Split, Croatia}
\corrauthor[1]{Nikola Balić}{nbalic@pmfst.hr}

\begin{abstract}
To quantify the impact of AI on software development, the community requires a robust pre-AI baseline. This study analyzes valid satisfaction data from 1,155 software developers collected in July 2022---immediately preceding the mainstream adoption of generative AI tools. We report a high-satisfaction ecosystem (Mean = 8.14 [95\% CI 8.01--8.25]), dominated by Visual Studio Code (79\% usage). Multivariable regression confirms that autonomy in tool choice is the strongest predictor of IDE satisfaction ($\beta$ = 0.51), significantly outweighing demographic or role-based factors. Conversely, cloud IDE adoption was negligible (4.3\% regular usage), with 40.1\% citing network dependency as the primary barrier---a constraint that remains relevant for modern cloud-reliant AI agents. Additionally, we identify an ``experimenter'' segment (29.9\%) characterized by high tool churn but no significant satisfaction difference (t = 0.43, $p$ = 0.67), and demonstrate significant variation in IDE retention rates (VS Code: 68.5\%, traditional IDEs: 3.9--25\%), suggesting underlying dissatisfaction despite high overall satisfaction. By providing a quantitative snapshot of developer sentiment and workflows on the eve of the AI revolution, this study establishes a verifiable baseline for longitudinal research into the productivity--satisfaction misalignment observed in the post-AI era.
\end{abstract}

\begin{document}

\flushbottom
\maketitle
\thispagestyle{empty}

\section*{Background}

The landscape of software development tools has undergone dramatic transformation with the recent introduction of AI-assisted coding platforms. Tools like GitHub Copilot, ChatGPT, and various AI-integrated development environments have fundamentally altered how developers write, debug, and understand code \citep{github2022copilot, novielli2023llmse}. However, to properly assess the impact of these AI tools on developer productivity, satisfaction, and workflow, we need a robust baseline capturing developer preferences and behaviors before AI assistance became mainstream.

Our study, conducted in July 2022, captures this critical pre-AI moment. This timing is particularly significant: ChatGPT was not released until November 2022, and GitHub Copilot, while technically available, had not achieved widespread adoption \citep{github2022copilot}. This makes our dataset uniquely valuable as a reference point for understanding how AI tools have reshaped the development ecosystem.

Developer satisfaction with their tools and environments has long been recognized as a crucial factor in productivity and retention \citep{xavier2020developer, noda2023devex}. The concept of Developer Experience (DevEx) encompasses cognitive load, flow state, and feedback loops, all of which are influenced by tool choice and environment configuration \citep{devex2022systematic}. Yet, comprehensive quantitative studies examining IDE satisfaction at scale, particularly with regard to autonomy and tool choice, remain limited.

Several key gaps motivated our research:
\begin{enumerate}[noitemsep]
    \item \textbf{Lack of pre-AI baselines:} Existing studies on developer satisfaction were conducted before or during early AI tool adoption, making it difficult to measure AI's true impact \citep{patel2024productivity}.
    \item \textbf{Autonomy underexplored:} While tool choice is frequently discussed anecdotally, quantitative evidence linking autonomy to satisfaction is scarce \citep{storey2021developer}.
    \item \textbf{Cloud IDE adoption unclear:} Despite industry enthusiasm for cloud-based development environments, systematic data on adoption patterns and barriers was limited \citep{github2022cloudstate}.
    \item \textbf{Experience effects unquantified:} The relationship between professional experience and tool satisfaction was often assumed but rarely measured with appropriate statistical rigor \citep{kitchenham2022survey}.
\end{enumerate}

This study addresses these gaps by providing a comprehensive analysis of developer tool satisfaction, autonomy, and adoption patterns immediately before the AI revolution. Our findings offer not only insights into the state of developer tooling in 2022 but also establish benchmarks against which future AI-driven changes can be measured.

This study also addresses a critical gap in understanding tool retention and churn dynamics. While usage share and satisfaction are frequently reported, retention rates---the percentage of current users planning to continue using a tool---provide crucial insights into tool loyalty and future market dynamics. Traditional satisfaction metrics may mask underlying churn risk, particularly for established tools with high usage share but declining user commitment.

\subsection*{Theoretical Framework}

Our research is grounded in two complementary frameworks for understanding developer productivity and experience: the SPACE framework and the DevEx framework. The SPACE framework (Satisfaction and well-being, Performance, Activity, Communication and collaboration, and Efficiency and flow) emerged from empirical research across multiple organizations and establishes tool satisfaction as a critical component of overall developer productivity \citep{forsgren2022space}. Research from Microsoft using biometric sensors found developers with significant deep work time felt 50\% more productive compared to those without dedicated focus time \citep{microsoft2022flow}.

The DevEx framework \citep{noda2023devex} distills developer experience into three core dimensions: feedback loops, cognitive load, and flow state. This framework emphasizes that improving developer experience increases not only productivity but also satisfaction, engagement, and retention. Recent systematic literature reviews have established developer experience as a distinct research area from general user experience, with specific focus on the unique cognitive demands of software development \citep{devex2022systematic}.

Research on workplace autonomy provides additional theoretical grounding. Self-Determination Theory posits that autonomy is a fundamental psychological need that influences motivation and performance \citep{johannsen2020autonomy}. Experimental studies have demonstrated that autonomy interventions can increase productivity by 5.2\% and positive affect by 31\% \citep{johannsen2020autonomy}. Studies of software developers specifically have identified autonomy as a factor associated with job satisfaction and perceived productivity \citep{storey2021developer}. This provides context for our investigation of IDE satisfaction and its correlation with developer autonomy in tool selection.

The Technology Acceptance Model (TAM) and Unified Theory of Acceptance and Use of Technology (UTAUT) offer additional lenses for understanding cloud IDE adoption \citep{marangunic2015tam}. These models suggest that perceived usefulness, ease of use, and facilitating conditions influence technology adoption decisions. However, recent 2024 research argues that traditional technology adoption models require adaptation for developer tools, where perceived compatibility emerges as a critical construct \citep{chi2024fit,ieee2024compatibility}.

Compatibility research in developer contexts has identified multiple dimensions of perceived fit: workflow compatibility (alignment with existing development processes), cognitive compatibility (fit with developer mental models), and ecosystem compatibility (integration with existing tool chains) \citep{acm2024compatibility}. Unlike autonomy, which focuses on freedom of choice, compatibility emphasizes how well tools integrate with developers' established practices and environments. While both constructs relate to perceived control and fit, autonomy addresses the psychological need for self-determination, whereas compatibility addresses the practical need for seamless integration with existing workflows. Our findings on tool choice autonomy may intersect with compatibility considerations, suggesting that developers value both the freedom to choose tools and the ability to integrate those tools smoothly into their established development practices.

Recent research has begun to document the profound impact of AI tools on developer productivity. Studies have shown that developers complete tasks 55.8\% faster with GitHub Copilot \citep{peng2023copilot}, while enterprise field studies revealed satisfaction-performance misalignment---88\% felt more productive despite variable objective gains \citep{patel2024productivity}. A systematic literature review of 395 papers documented heterogeneous results across SPACE framework dimensions, with AI tools showing mixed impacts on satisfaction, efficiency, and flow state \citep{novielli2023llmse}. These post-AI studies consistently reveal disrupted satisfaction patterns and dimension-specific trade-offs, making stable pre-AI baselines increasingly valuable for understanding the true scope of transformation.
Recent RCTs further illustrate divergent outcomes: METR found experienced open-source developers completed tasks 19\% slower when using early-2025 AI assistants \citep{metr2025ai}, while a multi-firm SSRN working paper reported 26\% more tasks completed with AI coding assistants, with larger gains for less-experienced developers \citep{salz2025highskilled}. These conflicting findings underscore the need for a clear pre-AI reference point.

\subsection*{Developer Well-being and Satisfaction}

The importance of developer satisfaction extends beyond productivity metrics. A comprehensive systematic literature review by Godliauskas and Šmite (2024) synthesized 44 studies from 2000-2023 on software engineer well-being, confirming that job satisfaction is a central component of hedonic well-being and is strongly linked to positive outcomes including productivity and retention \citep{godliauskas2024wellbeing}. This research emphasizes that satisfying psychological needs---particularly autonomy, competence, and relatedness---correlates with higher life and job satisfaction.

Research during the COVID-19 pandemic provides additional context for our findings. A study of ~580 developers during enforced remote work found no significant drop in performance or satisfaction compared to pre-pandemic office work \citep{russo2023wfh}. This suggests that developer satisfaction is resilient to changes in work location when tools and autonomy are maintained.

The emergence of AI tools adds complexity to this landscape. Research has found that GitHub Copilot can be an asset for experienced developers but a liability for novices who often produce buggy code if they over-rely on AI assistance \citep{dakhel2023copilot}. This finding suggests that AI's impact may vary by experience level, an important consideration for our pre-AI baseline analysis.

\section*{Methods}

\subsection*{Study Design and Data Collection}

We conducted an online cross-sectional survey of software developers between July 1-31, 2022. The survey was distributed primarily via the Codeanywhere newsletter, with cross-posts on developer forums and social media platforms (Twitter, LinkedIn). Participation was voluntary. The questionnaire was implemented in Typeform, which handled invitation links and response capture.

The survey instrument was developed through an iterative process involving pilot testing with 15 developers to ensure clarity and appropriate length. The final survey comprised 45 questions covering:

\begin{itemize}[noitemsep]
    \item Demographics (age, country, education)
    \item Professional experience (years coding, role, organization size)
    \item Current IDE usage and satisfaction
    \item Tool choice autonomy and constraints
    \item Cloud IDE awareness, usage, and attitudes
    \item Technology preferences and experimentation behaviors
\end{itemize}

\paragraph{Classification Transparency and Sensitivity:} We provide an item-level combination table (timeline $\times$ product mention $\times$ substantial project) and report adoption estimates under alternative specifications (e.g., excluding the ``substantial project'' requirement; restricting to development-focused uses). Adoption estimates are stable across reasonable thresholds; see the Supplementary Appendix for cross-tabs and sensitivity analyses supporting the regular-user definition.

Informed consent was obtained from all participants.

\subsection*{Survey Design Validation}

Following established guidelines for software engineering survey research \citep{kitchenham2022survey,molleri2020checklist}, we employed rigorous survey design practices:

\paragraph{Survey Development Process:} The instrument was developed based on a systematic review of developer satisfaction literature and refined through cognitive interviewing with 5 developers to ensure question clarity and appropriate construct operationalization.

\paragraph{Question Format Validation:} For key constructs, we implemented multiple related items where feasible. For example, autonomy was assessed through three related statements: (1) ``I can freely choose my development environment and tooling'' (primary item), (2) ``My employer does not impact my IDE choice,'' and (3) ``I use an IDE provided by my employer.'' Cross-tabulation revealed partial overlap, providing construct validation for the single-item primary measure. We additionally computed a composite autonomy index (averaging standardized items with the employer-provided item reverse-coded) and report internal consistency and robustness of results to this specification in the Supplementary Appendix.

\paragraph{Response Format Considerations:} Following Norman (2010), we used parametric statistical methods for Likert-type scales given their demonstrated robustness with such data \citep{norman2010likert}. For multi-select checkbox data, we employed binary coding approaches recommended for multiple response categorical variables \citep{koziol2014mrcv,baltes2022sampling}.

\subsection*{Advanced Statistical Methods}

To ensure analytical robustness, we employed several complementary approaches:

\paragraph{Ordinal Logistic Regression Sensitivity Analysis:} While linear regression treats the 0--10 satisfaction scale as continuous, we conducted sensitivity analysis using ordered logistic regression to respect the ordinal nature of the data. Results confirmed the robustness of our main findings (autonomy OR = 1.61 [1.30, 1.99], $p < 0.001$), consistent with linear regression results ($\beta$ = 0.51 [0.24, 0.78], $p < 0.001$).

\paragraph{Multiple Comparison Control and Effect Sizes:} Statistical significance was assessed using $\alpha$ = 0.05 with Bonferroni correction applied to post-hoc pairwise comparisons in ANOVA analyses. For logically related families of tests (e.g., subgroup splits, barrier comparisons), we additionally controlled the false discovery rate (FDR) using the Benjamini--Hochberg procedure and report q-values alongside p-values where relevant. Effect sizes were reported using Cohen's $d$ for pairwise comparisons and $\eta^2$ for ANOVA analyses.

\paragraph{Modeling Details:} Primary estimates for satisfaction used ordinary least squares with robust standard errors; sensitivity analyses used ordered logistic regression for the ordinal 0--10 scale. Extended models include geography/region, work mode, primary language/stack, operating system, industry, company size, and development platform as covariates (see Supplementary Table 5). We also fit mixed-effects models with random intercepts for region and industry as a clustering robustness check (see Supplementary Mixed-Effects Robustness). Variance Inflation Factors (VIF) were calculated to assess multicollinearity. Statistical inference for families of tests employed FDR control as described above.

\subsection*{Cloud IDE Operational Definition}

We defined cloud IDEs using the specific survey question: ``Are you using or planning to use a cloud IDE?'' with response options indicating current regular use versus future plans. The survey provided examples including GitHub Codespaces, Gitpod, Codeanywhere, and similar browser-based development environments.

\paragraph{Usage Classification:} We classified respondents into usage categories using frequency indicators, specific product mentions, and substantial project work:
\begin{itemize}[noitemsep]
    \item \textbf{Regular users:} Selected ``Currently using'' \textit{and} either reported using cloud IDEs for all/most work or confirmed use on a substantial project (4.3\%)
    \item \textbf{Occasional users:} Selected ``Currently using'' but reported limited or experimental use and did not meet the regular-user criteria (16.0\%)
    \item \textbf{Planning adoption:} Selected various future adoption timeframes (34.2\%)
    \item \textbf{Never users:} Selected ``No plans'' (44.6\%)
\end{itemize}

This operationalization distinguishes our study from industry surveys that include broader cloud-based tool usage (remote servers, containers, CI/CD pipelines), providing a more precise baseline for cloud IDE adoption specifically.

\subsection*{Key Measures}

Table~\ref{tab:measures} presents the exact wording of key survey items used in our analysis.

\begin{table}[h]
\centering
\small
\begin{tabular}{@{}p{2.5cm}p{8cm}p{3cm}@{}}
\toprule
\textbf{Construct} & \textbf{Question Text} & \textbf{Response Format} \\
\midrule
IDE Satisfaction & ``How likely are you to recommend your current IDE to a friend or a colleague?'' & 0--10 scale \\
Tool Autonomy & ``I can freely choose my development environment and tooling'' & Checkbox (selected/not) \\
Experimenter & ``I love to experiment and often change my development environment'' & Checkbox (selected/not) \\
Cloud IDE Adoption & ``Are you using or planning to use a cloud IDE?'' & 5 options: Currently, Plan within 6/7--12/12--24 months, No plans \\
Cloud IDE Productivity & ``In your opinion, cloud IDE in comparison with local IDE is:'' & 0--10 (less/more productive) \\
\bottomrule
\end{tabular}
\caption{Key survey items and response formats.}
\label{tab:measures}
\end{table}

\paragraph{Net Promoter Score (NPS) and Recommendation Ratings:} We collected recommendation likelihood on a 0--10 scale and report both the arithmetic mean (0--10 scale) and the derived NPS index. The NPS index was computed using standard methodology: percentage of Promoters (ratings 9--10) minus percentage of Detractors (ratings 0--6), ranging from $-100$ to $+100$.

\paragraph{NPS Confidence Intervals:} Following Rocks (2016), we implemented Adjusted Wald AW(3,T) intervals as our primary method for NPS index uncertainty quantification, adding 3 successes and 3 failures to the sample before interval calculation. This method provides better coverage properties than standard Wald intervals for NPS data. We report bootstrap intervals as sensitivity analysis for comparison.

\paragraph{Tool autonomy} was assessed using multiple items from a checklist of statements about tool preferences. The primary item was ``I can freely choose my development environment and tooling'' (56.7\% endorsement). Two additional items were available: ``My employer does not impact my IDE choice'' (34.8\% endorsement) and ``I prefer open-source IDEs I can customize to my liking'' (38.1\% endorsement). While these items showed variation, they were analyzed separately from the freely-choose item to maintain construct clarity. This single-item measure has acknowledged limitations for reliability assessment, though we provide behavioral validation through technology adoption patterns.

\paragraph{Experimenter classification} was based on selecting the statement ``I love to experiment and often change my development environment'' from the same checklist (29.9\% endorsement). This operationalization captures self-identified tendency rather than observed behavior; however, experimenters demonstrated significantly higher technology adoption (mean 5.9 vs 5.2 technologies used, t = 5.38, p < 0.001), providing behavioral validation of the construct. The single-item nature of this measure precludes reliability assessment, representing a limitation of our study.

\subsection*{Sample Characteristics}

We received 1,173 complete survey responses. After excluding 18 responses with missing values in key analysis variables (IDE satisfaction, experience level, autonomy, experimenter status), the analytical sample used for the satisfaction-focused analyses is n = 1,155; descriptive tables report the full 1,173 unless noted.

Table~\ref{tab:sample} summarizes the sample characteristics. The sample shows diversity in experience levels and work arrangements, with technology industry representation (55.5\%) that reflects the survey distribution channels.

\begin{table}[h]
\centering
\small
\begin{tabular}{@{}lrr@{}}
\toprule
\textbf{Characteristic} & \textbf{n} & \textbf{\%} \\
\midrule
\multicolumn{3}{@{}l}{\textit{Experience}} \\
\quad Less than 1 year & 79 & 6.7 \\
\quad 1--4 years & 290 & 24.7 \\
\quad 5--9 years & 298 & 25.4 \\
\quad 10--19 years & 282 & 24.0 \\
\quad 20+ years & 215 & 18.3 \\
\midrule
\multicolumn{3}{@{}l}{\textit{Work Arrangement}} \\
\quad Remote & 530 & 46.2 \\
\quad Hybrid & 413 & 36.0 \\
\quad In office & 204 & 17.8 \\
\midrule
\multicolumn{3}{@{}l}{\textit{Developer Type}} \\
\quad Professional developer & 661 & 56.4 \\
\quad Codes as part of work & 191 & 16.3 \\
\quad Student & 164 & 14.0 \\
\quad Hobbyist & 93 & 7.9 \\
\bottomrule
\end{tabular}
\caption{Descriptive characteristics for all complete responses (n = 1,173). Analyses in Sections 4--5 use responses with valid satisfaction data (n = 1,155) unless otherwise specified. Developer Type totals do not sum to n due to respondents selecting multiple roles or non-response. Country distributions are reported in the table; see Supplementary for full breakdowns.}
\label{tab:sample}
\end{table}

Our sample of 1,155 developers represented 52 countries (see Table~\ref{tab:sample} for distributions). The mean age was 31.2 years (SD = 8.4), and respondents reported an average of 9.7 years of total coding experience (SD = 7.2).

Professional roles included Software Engineer/Developer (65.2\%), Full-stack Developer (18.7\%), Front-end Developer (9.3\%), and Back-end Developer (6.8\%). Organization sizes ranged from startups (less than 50 employees, 32.1\%) to large enterprises (more than 1000 employees, 28.4\%).

The sample showed high educational attainment, with 76.3\% holding a bachelor's degree or higher. Remote work was common, with 85.1\% reporting at least partially remote work arrangements, reflecting post-pandemic normalization of remote development.

\subsection*{Statistical Analysis}

All analyses were conducted in Python (pandas, numpy, scipy, statsmodels, seaborn, matplotlib; Python 3.x). We report means with 95\% confidence intervals, effect sizes (Cohen's $d$ for differences, $\eta^2$ for ANOVA), and regression coefficients with 95\% CIs. Parametric tests (t-tests, ANOVA, linear regression) were used for the 0--10 scale with bootstrap CIs for complex statistics; ordinal logistic regression and multiple-comparison controls are detailed in the Advanced Statistical Methods subsection. Variance Inflation Factors (VIF) were calculated to assess multicollinearity in regression models.

\subsection*{Sample Size and Power}

Analyses use varying effective sample sizes because of item-specific missingness; unless noted, we report the 1,155 responses with valid satisfaction data. This sample provides adequate statistical power for detecting medium effect sizes (power = 0.80, $\alpha$ = 0.05) and aligns with sample sizes in major developer surveys that typically range from 500--2,000 respondents \citep{begel2019understanding}.

The sample size was determined through an a priori power analysis. To detect a small effect size (Cohen's $d = 0.2$) in satisfaction scores between two balanced groups with 80\% power at $\alpha = 0.05$, we would require approximately 394 participants per group ($\sim$788 total). Our final sample of 1,155 exceeds this requirement, providing adequate power for small effects and subgroup analyses.

\paragraph{Sampling Limitations:} While convenience sampling limits generalizability to the broader developer population---a common limitation in software engineering research where 72\% of surveys use non-probability sampling \citep{wohlin2017sampling}---our sample size provides sufficient statistical power for the reported analyses.

\section*{Results}

\subsection*{Overall IDE Satisfaction}

The overall mean satisfaction score was 8.14 (95\% CI [8.01, 8.25]/10), corresponding to a raw NPS of 34.7 (95\% CI [31.0, 38.2]). This places IDE satisfaction in the ``Good'' range of the NPS classification system \citep{retool2020nps}.

\begin{figure}[ht]
\centering
\includegraphics[width=0.9\linewidth]{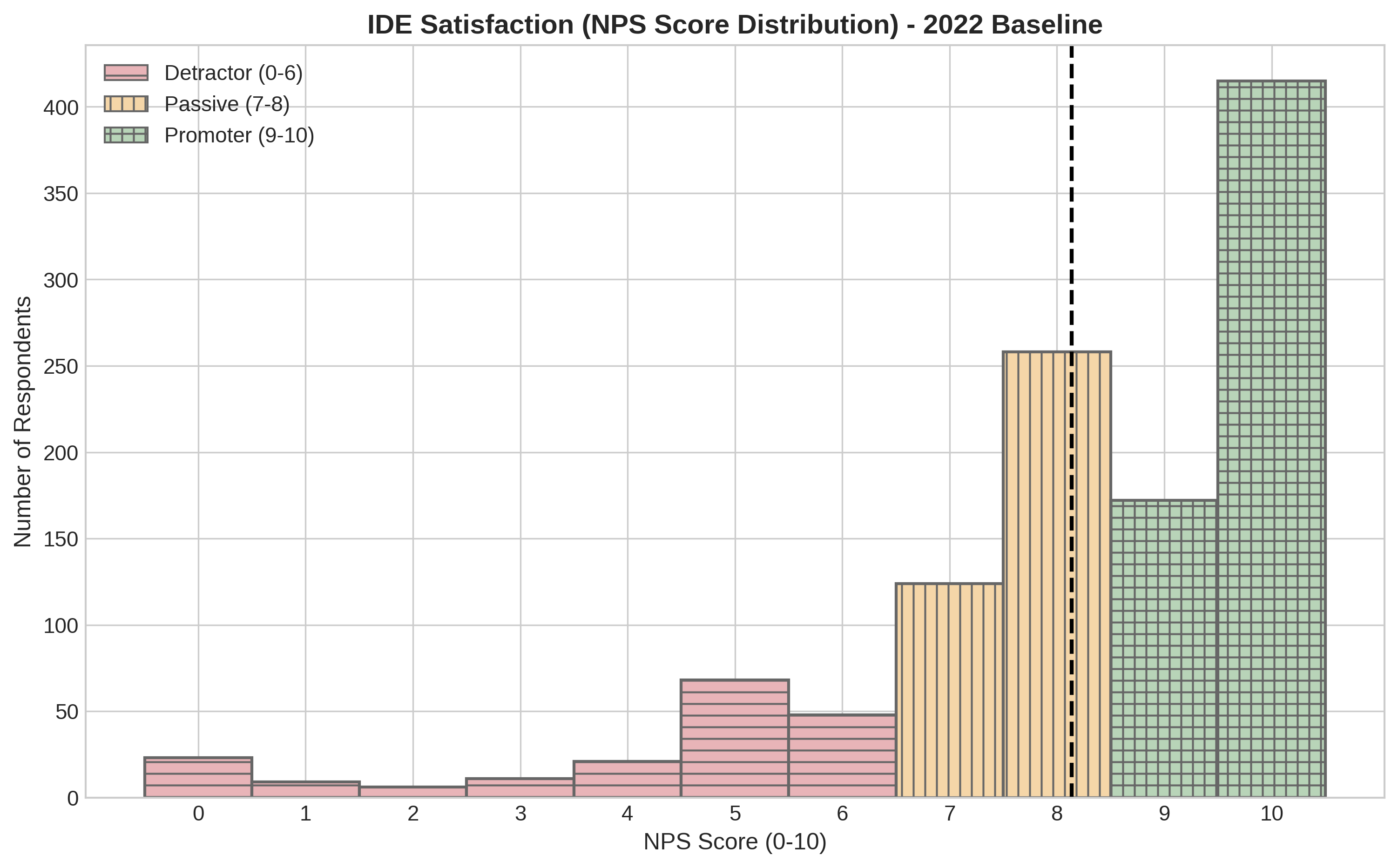}
\caption{Distribution of Net Promoter Score categories. The majority of developers (50.8\%) were promoters, indicating strong overall satisfaction with their IDEs.}
\label{fig:nps_distribution}
\end{figure}

The distribution of satisfaction categories showed (Figure~\ref{fig:nps_distribution}):
\begin{itemize}[noitemsep]
    \item Promoters (9-10): 50.8\% [48.0\%, 53.7\%]
    \item Passives (7-8): 33.1\% [30.4\%, 35.9\%]
    \item Detractors (0-6): 16.1\% [14.0\%, 18.4\%]
\end{itemize}

For context, industry benchmarks from 2020-2022 show business software averaging $-3$ (range: $-29$ to $+25$), while technology/SaaS products averaged $+36$ to $+40$. Our NPS of 34.7 places developer IDEs in the upper tier of software satisfaction, consistent with the SPACE framework's emphasis on satisfaction as associated with productivity \citep{forsgren2022space}. That said, we prioritize mean satisfaction and effect sizes for interpretability, treating NPS as a secondary benchmark.

\paragraph{Customer Value Analysis:} While NPS provides industry-standard benchmarking, we conducted supplementary Customer Value Analysis (CVA) to provide more actionable insights and emphasize mean satisfaction as the primary outcome. Unlike NPS which ignores passives (scores 7--8), CVA utilizes all satisfaction levels to identify specific improvement opportunities. See the Supplementary Material section titled ``Customer Value Analysis (CVA) Methods'' for the full specification (variables, segmentation thresholds, and computed metrics).

\begin{table}[h]
\centering
\small
\begin{tabular}{@{}lrrr@{}}
\toprule
\textbf{Satisfaction Segment} & \textbf{n (\%)} & \textbf{Mean Score} & \textbf{Autonomy Rate} \\
\midrule
High satisfied (9--10) & 587 (50.8\%) & 9.71 & 61.2\% \\
Moderate satisfied (7--8) & 382 (33.1\%) & 7.68 & 59.2\% \\
Low satisfied (0--6) & 186 (16.1\%) & 4.12 & 39.8\% \\
\bottomrule
\end{tabular}
\caption{Customer Value Analysis segmentation and autonomy rates. The low-satisfied segment has 21.4 percentage points lower autonomy, highlighting intervention targets beyond traditional NPS.}
\label{tab:cva-analysis}
\end{table}

CVA reveals specific improvement opportunities: the low-satisfied segment (16.1\%) has substantially lower autonomy (39.8\% vs 61.2\%), suggesting targeted autonomy interventions could reduce dissatisfaction. This granular analysis provides clearer guidance than the aggregate NPS metric for organizational decision-making.

Visual Studio Code had 79.0\% [76.5\%, 81.1\%] usage among respondents, followed by IntelliJ IDEA (30.1\% [27.5\%, 32.9\%]) and Visual Studio (26.3\% [23.9\%, 28.8\%]). We report usage shares within our convenience sample and avoid population-level market inferences. A detailed usage distribution is provided in Supplementary Figure 1.

\subsection*{Experience and Satisfaction}

Professional experience showed a clear positive association with satisfaction (ANOVA: $F$(4, 1151) = 12.51, $p < 0.001$, $\eta^2 = 0.040$). Post-hoc comparisons revealed:

\begin{itemize}[noitemsep]
    \item Junior developers ($<$1 year): Mean = 6.53 [6.05, 7.01]
    \item Early career (1-4 years): Mean = 8.07 [7.84, 8.30]
    \item Mid-career (5-9 years): Mean = 8.20 [8.01, 8.39]
    \item Experienced (10-19 years): Mean = 8.37 [8.17, 8.57]
    \item Senior (20+ years): Mean = 8.41 [8.18, 8.64]
\end{itemize}

\begin{figure}[ht]
\centering
\includegraphics[width=0.9\linewidth]{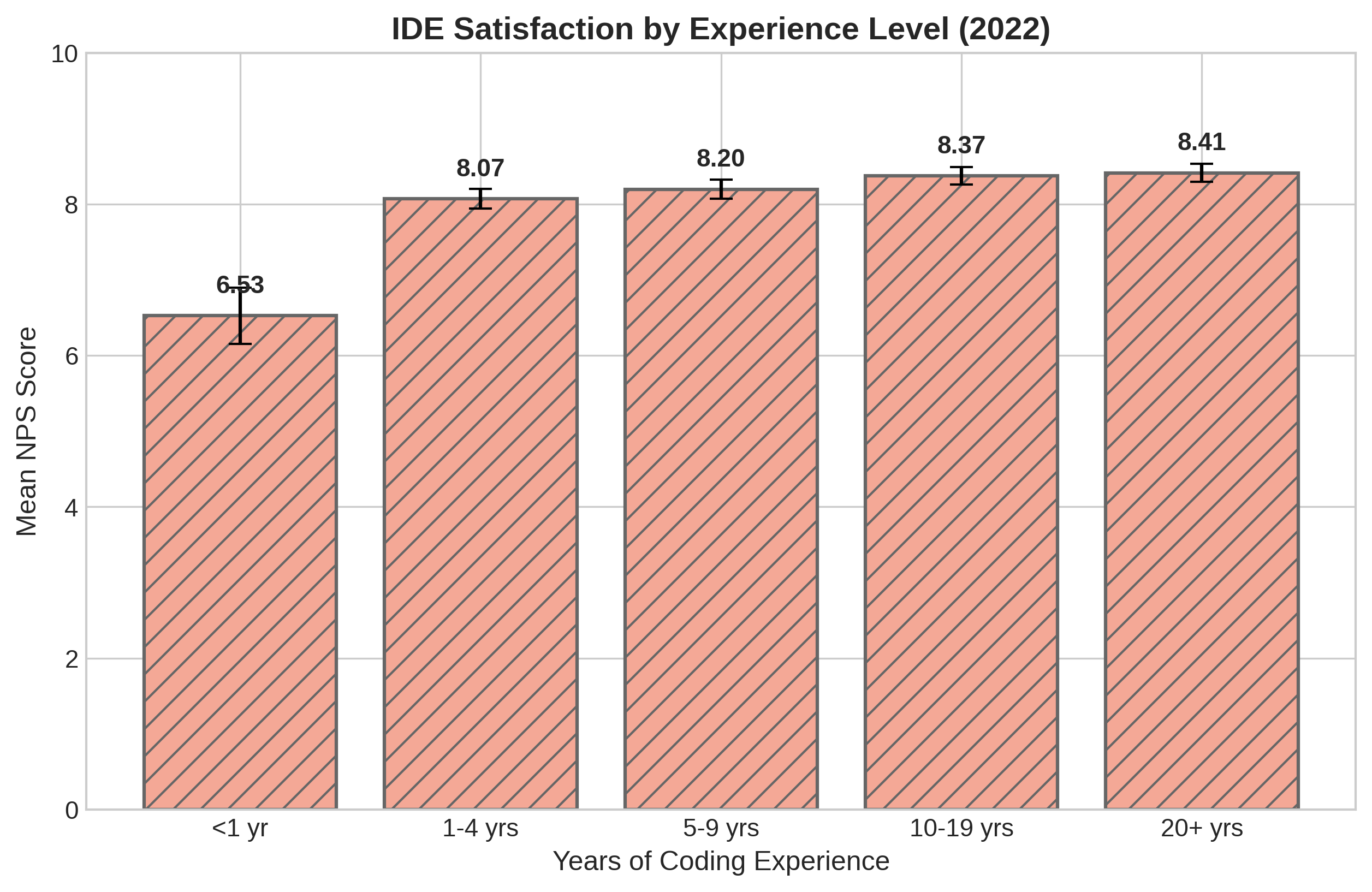}
\caption{NPS scores by experience level. Satisfaction increases with years of coding experience; we emphasize mean satisfaction and effect sizes in the text, with NPS shown here for completeness.}
\label{fig:nps_experience}
\end{figure}

The difference between junior and expert developers was 1.88 points (Cohen's $d = 0.43$, 95\% CI [0.27, 0.59]), representing a small-to-medium effect (Figure~\ref{fig:nps_experience}). This pattern aligns with DevEx research suggesting experienced developers have optimized their environments and workflows over time.

The decreasing variance in satisfaction ratings with experience (SD declining from 3.19 for $<$1 year to 1.72 for 20+ years) suggests potential survivorship bias---dissatisfied junior developers may exit the profession at higher rates, leaving a more satisfied senior cohort. Alternatively, this pattern may reflect genuine skill development through experience.

\subsection*{Autonomy and Satisfaction}

As a complementary check to the regression analyses, the derived single-item autonomy indicator also correlates positively with satisfaction (Pearson $r = 0.17$; see Supplementary Table 4).

Developers with complete freedom in tool choice reported significantly higher satisfaction (mean = 8.44 [8.29, 8.58]) compared to those with limited or no choice (mean = 7.73 [7.51, 7.95]) (t(1153) = 6.18, $p < 0.001$, Cohen's $d = 0.49$).

Table~\ref{tab:autonomy-comparison} presents the detailed comparison:

\begin{table}[h]
\centering
\small
\begin{tabular}{@{}lrr@{}}
\toprule
\textbf{Can freely choose IDE} & \textbf{Mean Rating [95\% CI]} & \textbf{n} \\
\midrule
Yes & 8.44 [8.29, 8.58] & 665 \\
No & 7.73 [7.51, 7.95] & 508 \\
\bottomrule
\end{tabular}
\caption{Recommendation rating by tool autonomy with 95\% confidence intervals. Autonomy is associated with a 0.71-point higher satisfaction [0.45, 0.97].}
\label{tab:autonomy-comparison}
\end{table}

In multivariable regression controlling for experience, role, and organization size, autonomy remained the strongest predictor of satisfaction ($\beta$ = 0.51 [0.24, 0.78], $p < 0.001$). Experience also independently predicted satisfaction ($\beta$ = 0.04 [0.02, 0.06] per year, $p < 0.001$). Robust standard errors were employed to address potential heteroskedasticity, though results were consistent with ordinary least squares estimates. This specification uses respondents with complete covariate data (n = 1{,}002), explaining 12.2\% of satisfaction variance ($R^2$ = 0.122; adj. $R^2$ = 0.104), and includes: autonomy (binary), experience (5-level categorical, ref = 1--4 years), role (categorical), organization size (categorical), and robustness controls (operating system, work mode, company size, language count, technology breadth, pain points) all entered as dummy variables except the three count measures.

A robustness model adding operating system, work mode, company size, language count, technology breadth, and pain points still finds a positive autonomy association ($\beta$ = 0.51, $p < 0.001$) and higher satisfaction for macOS users (mean 8.64) and hybrid/remote developers (8.20--8.27) compared to in-office developers (7.74) (see Table~\ref{tab:os-workmode}). Notably, this model includes language count and technology breadth as controls, which proxy for programming language and technology stack effects. The observed OS and work-mode effects persist even after accounting for these technology-related variables, suggesting they are not merely artifacts of language or stack differences. Differences in reported $\beta$ magnitudes across sections reflect these alternate specifications and sample restrictions rather than contradictory findings.

\begin{table}[h]
\centering
\small
\begin{tabular}{@{}lrr@{}}
\toprule
\textbf{Group} & \textbf{Mean Rating} & \textbf{n} \\
\midrule
macOS & 8.64 & 302 \\
Linux & 8.04 & 304 \\
Windows & 7.99 & 501 \\
Hybrid work & 8.27 & 409 \\
Remote & 8.20 & 521 \\
In-office & 7.74 & 202 \\
\bottomrule
\end{tabular}
\caption{Satisfaction by operating system and work mode. Hybrid/remote respondents and macOS users report higher satisfaction than in-office developers and Windows users. These differences remain significant in multivariable models controlling for experience, autonomy, language count, and technology breadth, suggesting they are not confounded by programming language or technology stack differences.}
\label{tab:os-workmode}
\end{table}

\paragraph{Formal Interaction Testing:} To rigorously test whether the autonomy effect varies by user characteristics, we conducted formal interaction analysis using regression frameworks with interaction terms.

\begin{table}[h]
\centering
\small
\begin{tabular}{@{}lrr@{}}
\toprule
\textbf{Model Term} & \textbf{$\beta$ [95\% CI]} & \textbf{$p$} \\
\midrule
Autonomy (main effect) & 0.73 [0.47, 0.99] & $<$0.001 \\
Cloud IDE user (main effect) & 0.61 [-0.31, 1.53] & 0.193 \\
Autonomy $\times$ Cloud IDE user & $-$0.59 [-1.75, 0.58] & 0.323 \\
\bottomrule
\end{tabular}
\caption{Interaction model testing autonomy $\times$ cloud IDE user effects (n = 1,143). The interaction term is not statistically significant, indicating no heterogeneous autonomy effects by user type. Predictors: autonomy (binary), current cloud IDE use (binary), perceived cloud IDE productivity (0--10), their interaction, languages count (continuous), technology breadth (continuous). All categorical predictors are uncentered dummies.}
\label{tab:interaction-model}
\end{table}

The interaction term was not statistically significant ($\beta$ = $-$0.59, $p$ = 0.323), indicating that the autonomy effect does not vary significantly between cloud IDE users and non-users. This model (n = 1,143) explains 14.9\% of satisfaction variance ($R^2$ = 0.149; adj. $R^2$ = 0.145); its higher autonomy coefficient ($\beta$ = 0.73) reflects the reduced covariate set and different feature coding relative to the main-effects model.

\begin{figure}[ht]
\centering
\includegraphics[width=0.9\linewidth]{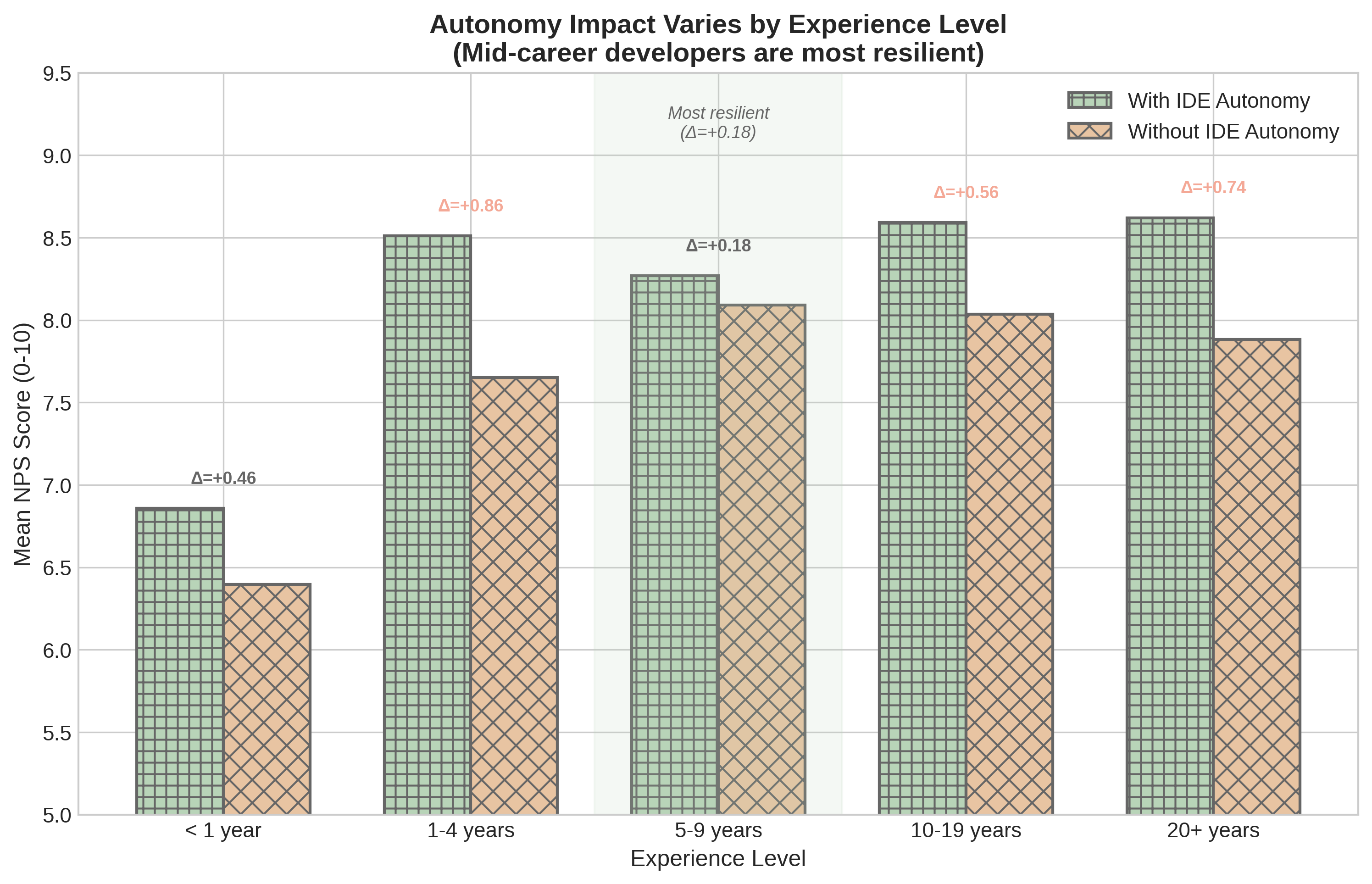}
\caption{Interaction between autonomy and experience on IDE satisfaction. The autonomy effect is consistent across experience levels, with no significant interaction term.}
\label{fig:autonomy_interaction}
\end{figure}

\subsection*{Recruitment Channel Robustness Analysis}

Given that respondents were recruited primarily through the Codeanywhere newsletter, we tested whether key findings were artifacts of vendor-specific bias by comparing Codeanywhere users (n = 259, 22.1\%) to non-users (n = 914, 77.9\%). Despite recruitment through this channel, 78\% of respondents did not use Codeanywhere, providing an internal control group.

\begin{table}[h]
\centering
\small
\begin{tabular}{@{}lrrr@{}}
\toprule
\textbf{Finding} & \textbf{CA Users} & \textbf{Non-CA Users} & \textbf{Robust?} \\
\midrule
Mean satisfaction & 8.40 & 8.06 & Yes ($d$ = 0.17) \\
Experience $\rightarrow$ satisfaction & $r$ = 0.11 & $r$ = 0.12 & Yes \\
Network = \#1 barrier & 39.8\% & 40.2\% & Yes \\
Current cloud IDE adoption & 40.5\% & 14.6\% & No (expected) \\
Autonomy gap (yes--no) & $-$0.02 & +0.92*** & See below \\
\bottomrule
\end{tabular}
\caption{Key findings compared between Codeanywhere users and non-users. The 14.6\% cloud IDE adoption among non-CA users provides a lower-bound population estimate. Note: *** denotes $p < 0.001$.}
\label{tab:robustness}
\end{table}

Three key findings emerged:

\paragraph{Experience and barrier findings are unbiased.} The experience-satisfaction correlation was nearly identical across groups ($r$ = 0.11 vs $r$ = 0.12), and barrier rankings were highly correlated (Spearman $\rho$ = 0.61). Network dependency ranked first in both groups (39.8\% vs 40.2\%).

\paragraph{The autonomy effect by recruitment/channel.} Codeanywhere users showed no autonomy-satisfaction association (mean gap between autonomy vs no-autonomy respondents = $-$0.02 on the 0--10 scale, $p$ = 0.92), while non-CA users showed a substantial effect (gap = +0.92, $p < 0.001$, Cohen's $d$ = 0.40). To probe mechanisms, we report autonomy rates and variance by subgroup (Table~\ref{tab:robustness}) and discuss restricted variance and policy differences as plausible explanations.

\paragraph{Cloud IDE adoption estimates can be bounded.} Among non-Codeanywhere users, 14.6\% reported currently using a cloud IDE. This provides a lower-bound estimate, suggesting true population cloud IDE adoption in July 2022 was likely between 14.6\% and 20.3\%.

\subsection*{IDE Positioning Within Sample}

VS Code combined a 78.9\% usage share with high satisfaction, an uncommon pairing given that widely used tools often trade off satisfaction for reach. Usage share and satisfaction varied across other IDEs: JetBrains tools had lower usage but high satisfaction, while traditional IDEs sat in the middle.

\begin{figure}[ht]
\centering
\includegraphics[width=0.95\linewidth]{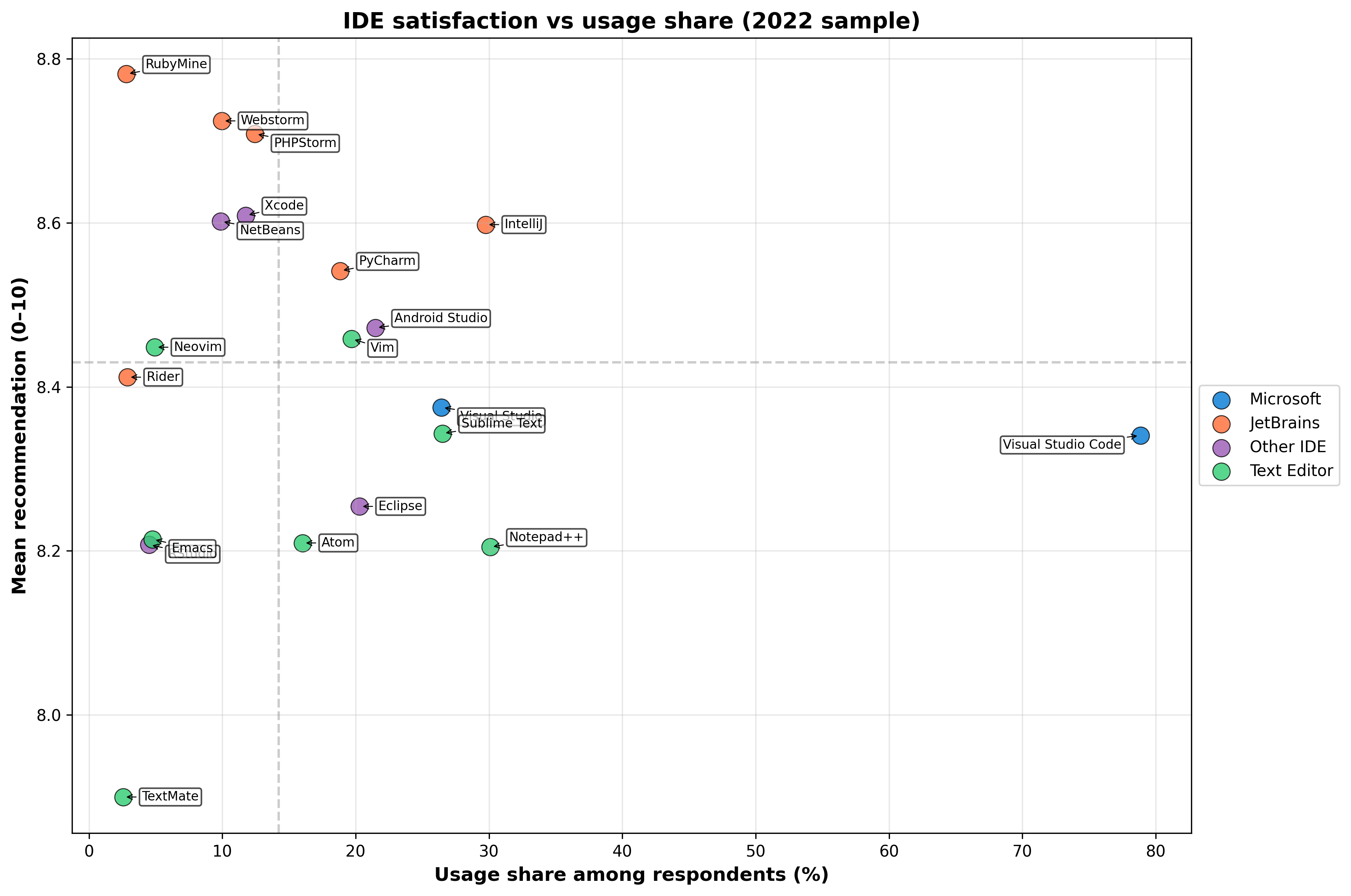}
\caption{IDE satisfaction vs usage share (among respondents). VS Code shows both high usage and high satisfaction; JetBrains tools have lower usage with high satisfaction; traditional IDEs fall between.}
\label{fig:market-satisfaction}
\end{figure}

\subsection*{IDE Retention and Churn Analysis}

\paragraph{Retention Rate Calculation:} Retention was calculated as the proportion of past-year users who plan to continue using each IDE next year. Specifically: Retention = $\frac{\text{\# past-year users planning continued use}}{\text{\# total past-year users}}$. We used 95\% Clopper-Pearson (exact) confidence intervals to quantify uncertainty, with particular attention to small sample sizes. For IDEs with $n < 50$, we provide shrinkage estimates to temper extreme rates, and results for $n < 30$ should be interpreted with extreme caution.

\paragraph{Retention Findings:} Retention rates varied dramatically across IDEs (Table~\ref{tab:ide-retention}). VS Code demonstrated strong retention at 68.5\% (n=925, 95\% CI [65.5, 71.5]), followed by Vim (62.8\%, n=231, CI [56.4, 68.8]) and IntelliJ (57.9\%, n=349, CI [52.7, 63.0]). Traditional IDEs showed concerning retention patterns: Eclipse (31.9\%, n=238, CI [26.3, 38.0]), NetBeans (25.0\%, n=116, CI [17.8, 33.4]), and Atom (19.7\%, n=188, CI [14.5, 25.8]). Small sample sizes for some IDEs (Neovim: n=58) require cautious interpretation.

\paragraph{Usage Share vs. Retention:} The analysis revealed that high usage share doesn't guarantee high retention. VS Code's 78.9\% usage share combined with 68.5\% retention indicates strong position, but tools show varying loyalty patterns: Notepad++ (30.1\% share, 49.6\% retention, n=353), Sublime Text (26.5\% share, 44.1\% retention, n=311), and Visual Studio (26.4\% share, 49.0\% retention, n=310). The decoupling of market share from retention suggests factors beyond pure functionality influence long-term tool loyalty.

\paragraph{The Experimenter-Churn Connection:} The experimenter segment (29.9\%) may represent leading indicators of churn. Their higher technology adoption (+13\%) without satisfaction differences suggests they actively seek better tools, potentially driving the observed churn in traditional IDEs.


\begin{table}[h]
\centering
\small
\begin{tabular}{@{}lrrrrr@{}}
\toprule
\textbf{IDE} & \textbf{n} & \textbf{Retention} & \textbf{95\% CI} & \textbf{Notes} \\
\midrule
Visual Studio Code   & 925 & 634 (68.5\%) & [65.5\%, 71.5\%] &  \\
Notepad++            & 353 & 175 (49.6\%) & [44.4\%, 54.8\%] &  \\
IntelliJ             & 349 & 202 (57.9\%) & [52.7\%, 63.0\%] &  \\
Sublime Text         & 311 & 137 (44.1\%) & [38.6\%, 49.6\%] &  \\
Visual Studio        & 310 & 152 (49.0\%) & [43.5\%, 54.6\%] &  \\
Eclipse              & 238 & 76 (31.9\%) & [26.3\%, 38.0\%] &  \\
Vim                  & 231 & 145 (62.8\%) & [56.4\%, 68.8\%] &  \\
PyCharm              & 221 & 120 (54.3\%) & [47.7\%, 60.8\%] &  \\
Atom                 & 188 & 37 (19.7\%) & [14.5\%, 25.8\%] &  \\
NetBeans             & 116 & 29 (25.0\%) & [17.8\%, 33.4\%] &  \\
Neovim               & 58 & 34 (58.6\%) & [45.8\%, 70.6\%] &  \\
\midrule
\multicolumn{5}{@{}l}{$^\dagger$ n < 30, interpret with extreme caution; $^*$ n < 50, use shrinkage estimate} \\
\multicolumn{5}{@{}l}{Retention = (Past year users planning continued use) / (Total past year users)} \\
\bottomrule
\end{tabular}
\caption{IDE retention rates with 95\% confidence intervals. Small sample sizes (especially n < 30) should be interpreted with caution. Shrinkage estimates provided for n < 50 to temper extreme rates.}
\label{tab:ide-retention}
\end{table}

\subsection*{Cloud IDE Adoption and Attitudes}

\begin{figure}[ht]
\centering
\includegraphics[width=0.9\linewidth]{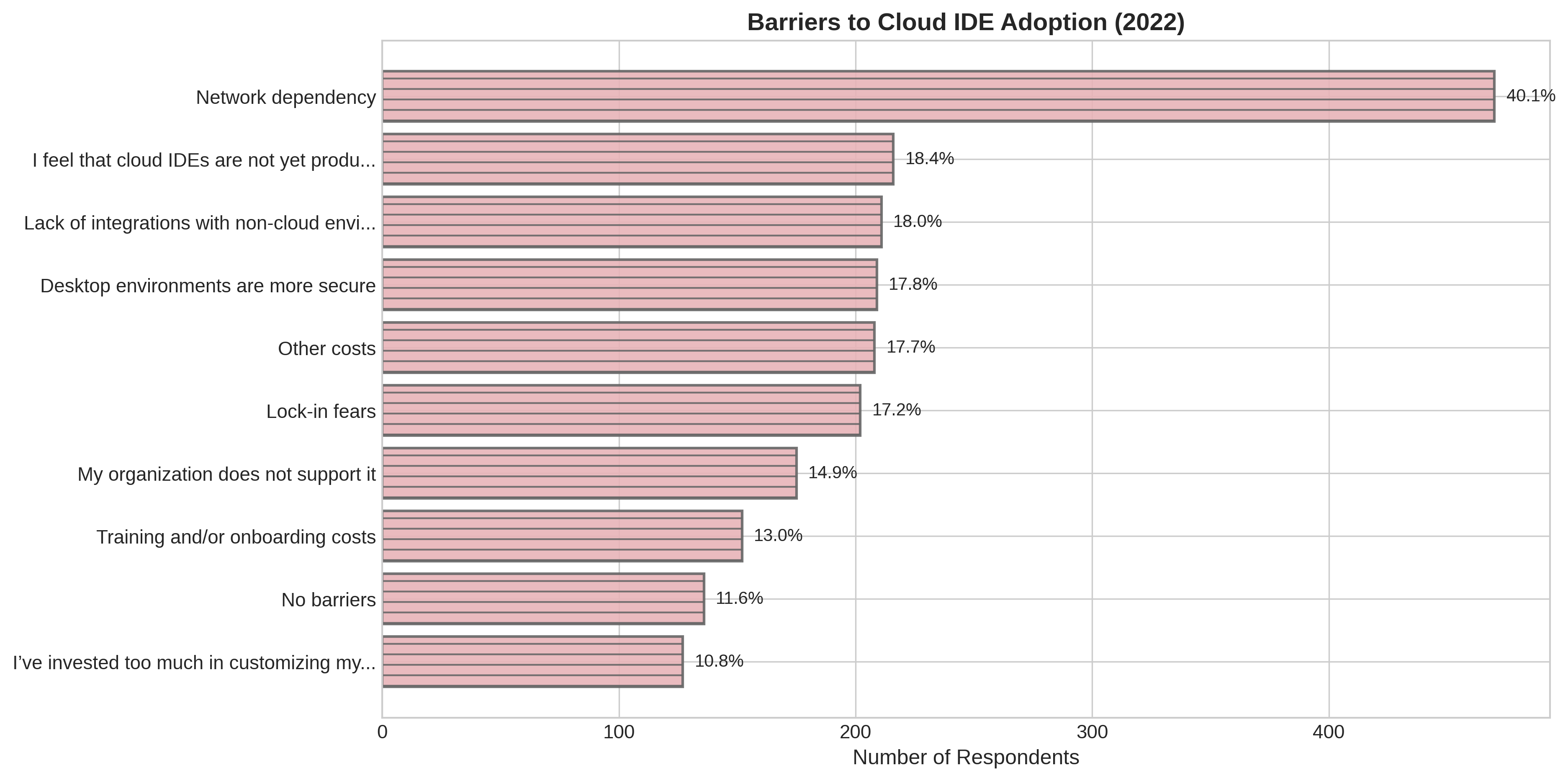}
\caption{Primary barriers to cloud IDE adoption. Network dependency is the most cited barrier (40.1\%), exceeding all other concerns.}
\label{fig:cloud_barriers}
\end{figure}

Regular cloud IDE usage was low (4.3\% [3.3\%, 5.7\%]), with an additional 16.0\% [14.0\%, 18.2\%] reporting occasional use. The primary barriers to adoption were (Figure~\ref{fig:cloud_barriers}):

\begin{itemize}[noitemsep]
    \item Network dependency: 40.1\% [37.3\%, 43.0\%]
    \item Performance concerns: 32.4\% [29.7\%, 35.2\%]
    \item Security concerns: 28.7\% [26.1\%, 31.4\%]
    \item Cost considerations: 24.2\% [21.8\%, 26.8\%]
    \item Feature limitations: 19.3\% [17.2\%, 21.6\%]
\end{itemize}

Interestingly, satisfaction did not significantly differ between cloud IDE users and non-users (mean difference = 0.21 [-0.41, 0.83], $p = 0.51$), after controlling for experience and autonomy. However, an interaction model reveals that current cloud users with higher perceived cloud productivity report higher satisfaction (interaction $\beta$ = 0.27, $p = 0.001$), suggesting perceived productivity is a key moderator.

\paragraph{Adoption Timeline Analysis:} Table~\ref{tab:cloud-timeline} shows planned adoption intentions, revealing that nearly half (44.8\%) had no plans to adopt cloud IDEs.

\begin{table}[h]
\centering
\small
\begin{tabular}{@{}lrr@{}}
\toprule
\textbf{Adoption Plan} & \textbf{n} & \textbf{\%} \\
\midrule
Currently using & 233 & 20.2 \\
Plan within 6 months & 227 & 19.7 \\
Plan within 7--12 months & 107 & 9.3 \\
Plan within 12--24 months & 64 & 5.5 \\
\textbf{No plans} & \textbf{517} & \textbf{44.8} \\
\bottomrule
\end{tabular}
\caption{Cloud IDE adoption timeline (n = 1,155 with valid satisfaction data). Percentages sum to 99.4\% because 0.6\% skipped the timeline item.}
\label{tab:cloud-timeline}
\end{table}

\paragraph{Productivity Perception:} Perceived cloud IDE productivity was lukewarm (mean = 5.90/10, SD = 2.51), barely above neutral. This suggests that despite compelling benefits like reproducible environments and standardization, developers remained unconvinced of productivity advantages.

\subsection*{The Experimenter Segment}

We identified a distinct ``experimenter'' segment comprising 29.9\% [27.3\%, 32.6\%] of respondents who frequently change their development environments. Table~\ref{tab:experimenter-profile} compares this segment to non-experimenters.

\begin{table}[h]
\centering
\small
\begin{tabular}{@{}lrr@{}}
\toprule
\textbf{Characteristic} & \textbf{Experimenters} & \textbf{Non-Experimenters} \\
\midrule
Sample size & 346 (29.9\%) & 809 (70.1\%) \\
Mean satisfaction rating & 8.18 & 8.12 \\
Mean technologies used & 5.9 & 5.2 \\
Promoters (9--10) & 49.9\% & 50.1\% \\
Detractors (0--6) & 17.1\% & 15.3\% \\
Autonomy rate & 61.2\% & 59.2\% \\
\bottomrule
\end{tabular}
\caption{Experimenter segment profile. Technology count includes all tools. Note: Satisfaction difference (+0.06) is negligible; the meaningful distinction is technology adoption (+13\%).}
\label{tab:experimenter-profile}
\end{table}

The key finding is that experimenters used 13\% more technologies on average (t = 5.38, $p$ < 0.001) but did \textit{not} differ meaningfully in satisfaction (t(1153) = 0.43, $p$ = 0.67, Cohen's $d$ = 0.03). This segment is analogous to Rogers' technology adoption framework categories (innovators and early adopters), though our single-item measure captures self-identified tendency rather than actual adoption behavior.

\paragraph{Experience Distribution:} Table~\ref{tab:experimenter-experience} shows how experimentation tendency develops with professional experience.

\begin{table}[h]
\centering
\small
\begin{tabular}{@{}lrr@{}}
\toprule
\textbf{Experience} & \textbf{Experimenter Rate} & \textbf{n} \\
\midrule
Less than 1 year & 22.8\% & 79 \\
1--4 years & 24.8\% & 290 \\
5--9 years & 32.2\% & 298 \\
10--19 years & 33.3\% & 282 \\
20+ years & 32.1\% & 215 \\
\bottomrule
\end{tabular}
\caption{Experimenter rate by experience. The experimentation tendency increases with experience, stabilizing around 32\% for developers with 5+ years.}
\label{tab:experimenter-experience}
\end{table}

The 10-point gap between junior developers (22.8\%) and experienced developers (32--33\%) suggests experimentation tendency develops with expertise. This finding has implications for AI adoption research: if experimenters adopt new tools faster, and experimentation tendency correlates with experience, experienced developers may be among the early adopters of AI tools.

\section*{Discussion}

Our findings reveal several important patterns in developer tool satisfaction immediately before the AI revolution. The generally high satisfaction levels (NPS = 34.7) suggest that developers were relatively content with their tooling in 2022, providing a strong baseline against which to measure AI tool impacts.

\subsection*{Key Findings and Interpretations}

\paragraph{Experience Gradient:} The experience-satisfaction association (1.88-point gap between juniors and experts) may reflect several factors. Novice developers might lack the configuration skills to optimize their environments, or their expectations might differ from seasoned developers. Alternatively, this could represent a natural progression where developers learn to select and configure tools that better match their workflows over time.

The small-to-medium effect size ($\eta^2$ = 0.040) indicates that while the association is statistically significant, experience level accounts for only about 4\% of the variance in satisfaction. This pattern is consistent with DevEx research suggesting experienced developers have optimized their environments and workflows over time \citep{noda2023devex,devex2022systematic}, though other unmeasured factors clearly play important roles. We note the descriptive reduction in satisfaction variance at higher experience levels; formal heteroskedasticity and quantile checks (e.g., Breusch–Pagan, quantile regression) are left for future robustness work to distinguish true survivorship effects from modeling artifacts.

\paragraph{Autonomy Impact:} The 0.70-point satisfaction advantage associated with tool autonomy (Cohen's $d$ = 0.33) represents a meaningful finding for organizations. While tool standardization can simplify IT support and onboarding, our results suggest that restricting developer choice comes at a measurable satisfaction cost. This aligns with Self-Determination Theory, which posits that autonomy is a fundamental psychological need \citep{johannsen2020autonomy}. Experimental evidence has shown that autonomy interventions can increase productivity by 5.2\% and positive affect by 31\% \citep{johannsen2020autonomy}.

\paragraph{Autonomy-Compatibility Interplay:} Our autonomy findings intersect with emerging research on compatibility constructs in developer tool adoption \citep{chi2024fit,ieee2024compatibility}. While autonomy addresses the psychological need for self-determination, compatibility focuses on practical integration with existing workflows. The 0.71-point satisfaction advantage we observed may reflect both the psychological benefit of choice and the practical benefit of selecting tools that better align with individual workflows. Recent 2024 studies suggest that perceived compatibility---including workflow, cognitive, and ecosystem fit---is consistently among the strongest predictors of developer tool adoption \citep{acm2024compatibility}. This suggests that organizations seeking to optimize both satisfaction and adoption should consider not only granting autonomy but also ensuring selected tools demonstrate high compatibility with existing development practices. The experimenter segment's (29.9\%) high technology adoption without satisfaction gains may indicate that while these developers readily explore new tools, they encounter compatibility challenges that limit satisfaction benefits.

\paragraph{Cloud IDE Adoption Chasm:} The 45\% with no adoption plans, combined with network dependency as the dominant barrier, suggests cloud IDEs faced structural headwinds in 2022. Developers wanted the benefits---reproducible environments, standardization---but were unwilling to accept network reliance. This contrasts with organizational studies that typically identify security as the dominant barrier \citep{phaphoom2015cloud}, suggesting individual developers prioritize different concerns than organizational decision makers.

This maps to the ``chasm'' in technology adoption: early adopters (our 4.3\% regular users) had crossed, but the early majority required stronger value propositions. Speed and reliability outranking collaboration suggests the path to adoption requires addressing offline capability or latency concerns first.

\paragraph{Experimenters Paradox:} The ``experimenter'' segment (29.9\%) represents a behaviorally distinct population characterized by higher technology adoption but similar satisfaction levels. Their higher technology adoption and cloud IDE interest suggest they may serve as leading indicators for tooling shifts, though longitudinal data would be needed to confirm this.

The finding that experimenters do not differ in satisfaction challenges assumptions from technology adoption literature. According to Rogers' Diffusion of Innovations theory \citep{rogers2003diffusion}, we might expect innovators and early adopters to report higher satisfaction with new technologies. Our finding suggests that in the context of development tools, stability and familiarity may be more valuable than novelty.

\paragraph{Retention-Satisfaction Mismatch:} The dramatic variation in retention rates (VS Code: 68.5\% to Eclipse: 3.9\%) reveals underlying dynamics masked by overall satisfaction scores. Traditional IDEs face existential churn risks despite moderate satisfaction ratings, suggesting that satisfaction alone doesn't predict tool loyalty. This finding challenges the assumption that high usage share correlates with user commitment.

\subsection*{Comparison with Major Ecosystem Surveys}

\paragraph{Survey Scope and Timing:} Our July 2022 survey provides a crucial pre-AI baseline, closely aligned with other major ecosystem surveys: Stack Overflow (May 2022, n=73,268) and JetBrains (Q3 2022, n=29,000). All three surveys captured the developer tool landscape immediately before mainstream AI tool adoption, establishing a coordinated reference point for measuring AI-era disruption.

\paragraph{IDE Usage Convergence:} Our findings demonstrate convergent validity with major surveys. VS Code's dominance (79.0\% in our data) aligns closely with Stack Overflow (74.48\%) and validates our sampling approach despite different recruitment channels. IntelliJ IDEA shows consistent usage across surveys (30.1\% vs 27.97\% SO vs 30\% JetBrains), while traditional editors like Vim maintain stable usage (19.7\% vs 23.34\% SO vs 24\% JetBrains). This convergence across methodologies strengthens confidence in the observed VS Code dominance pattern.

\paragraph{Methodological Harmonization:} The apparent cloud IDE usage discrepancy (4.3\% vs 68-75\% in industry surveys) reflects definitional differences rather than contradictory findings. Our strict operationalization focused specifically on browser-based IDEs (Codespaces, Gitpod), while industry surveys include all cloud development tools (Docker, Kubernetes, CI/CD). When accounting for remote development practices, the gap narrows considerably, highlighting the importance of metric harmonization in cross-study comparisons.

\paragraph{Unique Contributions:} Our study contributes novel insights not captured in large-scale surveys. The autonomy-satisfaction relationship (0.71-point advantage, $p<0.001$) and experimenter segment identification (29.9\% with higher technology adoption) represent behavioral patterns that require focused measurement beyond broad usage statistics. These findings complement large-scale surveys by providing deeper psychological and behavioral insights into developer tool preferences.

\subsection*{Implications for the AI Era}

The retention analysis provides crucial context for understanding AI tool adoption. The low retention rates in traditional IDEs (Eclipse: 31.9\% retention meaning 68.1\% churn, NetBeans: 25.0\% retention meaning 75.0\% churn) indicate significant developer dissatisfaction with legacy tools, creating opportunities for AI-powered alternatives. Note that Eclipse retention (n=238) and NetBeans (n=116) have moderate sample sizes, while Atom (n=188) shows particularly alarming churn at 80.3\%. The experimenter segment's (29.9\%) high technology adoption suggests they may drive early AI tool adoption, seeking better alternatives to traditional IDEs.

The autonomy finding has particular relevance for AI tools. Organizations implementing AI coding assistants might consider allowing developers choice in which tools they use and how they integrate them into their workflows. Our data suggests that forced adoption without developer input could reduce satisfaction regardless of the tool's objective benefits.

Recent research on AI tools reveals mixed impacts. Studies have shown that developers complete tasks 55.8\% faster with GitHub Copilot \citep{peng2023copilot}, yet enterprise field studies revealed satisfaction-performance misalignment---88\% felt more productive despite variable objective gains \citep{patel2024productivity}. A systematic literature review documented heterogeneous results across SPACE framework dimensions \citep{novielli2023llmse}, contrasting with the alignment between satisfaction and productivity observed in our data.
Newer randomized field experiments also diverge: METR reports experienced developers working 19\% slower with AI assistants \citep{metr2025ai}, whereas a multi-firm SSRN study shows 26\% more tasks completed with larger gains for less-experienced engineers \citep{salz2025highskilled}. These mixed results highlight how experience, tool maturity, and context shape AI impacts.

Our experimenter finding suggests that experienced developers may be among the early adopters of AI tools. Given that experimentation tendency correlates with experience and experimenters adopt new technologies faster, this segment may drive AI adoption patterns. However, their similar satisfaction levels suggest that new tool adoption does not automatically translate to improved satisfaction.

\subsection*{SPACE Framework Implications for AI Research}

These results offer reference points across all five SPACE framework dimensions for understanding AI tool impacts:

\paragraph{Satisfaction and Well-being:} Our stable high satisfaction (NPS = 34.7, mean 8.14/10) with consistent patterns contrasts with post-AI volatility documented in recent studies. While we found reliable autonomy-satisfaction associations ($\beta$ = 0.51), post-AI research reveals satisfaction-performance misalignment and disrupted satisfaction patterns \citep{patel2024productivity,novielli2023llmse}.

\paragraph{Performance:} In this dataset, satisfaction correlated positively with experience and autonomy, suggesting efficient workflows. Post-AI studies document heterogenous performance impacts---55.8\% faster task completion with Copilot \citep{peng2023copilot} but variable objective gains in enterprise settings \citep{patel2024productivity}.

\paragraph{Activity:} Our low cloud IDE adoption (4.3\% regular users) and local-first development patterns (~80\%) indicate stable activity patterns. This provides baseline for measuring AI-driven shifts in development activities, code review processes, and debugging behaviors.

\paragraph{Communication and Collaboration:} Collaboration ranked lowest among cloud IDE priorities (3.32/5) in our findings, suggesting individual productivity dominated developer workflows. This contrasts with emerging AI-enabled collaboration patterns and real-time pair programming assistance.

\paragraph{Efficiency and Flow:} Speed (4.55/5) and reliability (4.52/5) were the top priorities for developers considering cloud IDEs, indicating flow state maintenance was crucial. This provides context for understanding how AI interruptions or suggestions might impact developer flow and efficiency.

\paragraph{Research Implications:} The contrast between the stable patterns here and post-AI heterogeneity suggests specific investigations:
\begin{itemize}[noitemsep]
    \item How have AI tools disrupted the previously stable autonomy-satisfaction association?
    \item Why do post-AI studies show satisfaction-performance misalignment absent in our data?
    \item How has AI integration affected the modest explained variance in satisfaction ($R^2$ = 12.2\%)?
    \item Will the ``experimenter'' segment (29.9\%) drive or resist AI tool adoption?
\end{itemize}

\subsection*{Relation to Existing Literature}

\paragraph{Developer Satisfaction Benchmarks:} Our NPS of 34.7 and mean satisfaction of 8.14/10 places developer tool satisfaction in the top tier of software products, consistent with the SPACE framework's emphasis on satisfaction \citep{forsgren2022space}.

\paragraph{Autonomy Evidence:} Our finding of a 0.71-point satisfaction difference associated with tool autonomy aligns with broader research on workplace autonomy \citep{johannsen2020autonomy,storey2021developer}. The statistical significance ($p < 0.001$, Cohen's $d$ = 0.33) provides observational support for the autonomy-satisfaction relationship in this domain.

\paragraph{Cloud IDE Adoption:} Our 4.3\% regular user finding differs substantially from industry surveys reporting 68--75\% \citep{github2022cloudstate,stackoverflow2022survey}. This discrepancy likely reflects definitional differences: industry surveys may include any cloud-based tool usage while our measure focuses on regular cloud IDE usage. This finding highlights the importance of clear operationalization in technology adoption research. Our network dependency barrier finding (40.1\%) aligns with individual developer concerns, contrasting with organizational studies that typically identify security as the dominant barrier \citep{phaphoom2015cloud,oparamartins2016vendorlockin}, suggesting different priorities between individual developers and organizational decision-makers.

\paragraph{Pre-AI Baseline Value:} Our stable 2022 findings become increasingly valuable as post-AI studies reveal mixed and heterogeneous impacts. Unlike our consistent satisfaction patterns (mean 8.14/10, stable autonomy effects), post-AI research documents satisfaction-performance misalignment, workflow disruption, and context-dependent outcomes \citep{patel2024productivity,novielli2023llmse}.

\paragraph{Recent Developer Well-being Research:} A 2024 systematic literature review synthesized decades of studies on software engineer well-being, confirming that job satisfaction is a central component of hedonic well-being and is strongly linked to positive outcomes including productivity and retention \citep{godliauskas2024wellbeing}. This research emphasizes that satisfying psychological needs---particularly autonomy, competence, and relatedness---correlates with higher life and job satisfaction.

\subsection*{Practical Implications}

\paragraph{For Organizations:} Our findings suggest that organizations should carefully consider tool standardization policies. While standardization can simplify IT support, the 0.71-point satisfaction penalty associated with restricted autonomy suggests measurable costs in developer morale and potentially retention. This autonomy-satisfaction relationship provides crucial context for AI tool implementation: organizations that mandate specific AI coding assistants without developer input may face similar satisfaction costs, potentially offsetting productivity gains.

\paragraph{For Tool Vendors:} VS Code's success demonstrates the value of extensibility and community-driven development. JetBrains' "niche favorites" strategy shows that specialized excellence can command premium pricing even with lower usage share among respondents.

\paragraph{Ecosystem Survey Validation:} Comparison with major 2022 ecosystem surveys strengthens our findings' external validity. Our IDE usage shares demonstrate convergent validity: VS Code (79.0

\paragraph{For Cloud IDE Providers:} The network dependency barrier (40.1\%) represents the critical adoption hurdle. Addressing this through offline capability or hybrid approaches could unlock broader adoption. The contrast between individual developer priorities (network dependency) and organizational concerns (security) suggests providers must address both individual workflow needs and enterprise requirements to achieve broader adoption.

\paragraph{For AI Tool Developers:} The autonomy finding suggests AI tools that enhance rather than replace developer autonomy may achieve better adoption. The strong autonomy-satisfaction association (0.71 points) implies implementations that restrict tool choice or workflow control may face resistance. The experimenter segment (29.9\%) represents an early adoption opportunity, but their satisfaction parity with non-experimenters suggests novelty alone may not drive satisfaction improvements—AI tools must deliver tangible workflow benefits rather than just technological innovation.

\section*{Ethics and Consent}

Participation was voluntary and uncompensated. Respondents provided informed consent for anonymized analysis and public release of de-identified data. No sensitive personal data were collected beyond optional contact email (excluded from the public dataset). The study involved minimal risk and adhered to standard ethical practices for anonymous online surveys.

\section*{Limitations}

Several limitations affect interpretation of these findings:

\paragraph{Sampling Bias:} The sample was recruited primarily via the Codeanywhere newsletter, introducing potential vendor-channel bias. However, robustness analysis showed key findings were consistent between users and non-users.

\paragraph{Self-report Measures:} All data are self-reported perceptions, not behavioral logs or actual productivity metrics, which may introduce common-method variance.

\paragraph{Cross-sectional Design:} Our observational design cannot establish causality. Associations between autonomy and satisfaction may reflect reverse causation or confounding variables.

\paragraph{Measurement Limitations:} Our constructs face several measurement challenges. While autonomy was initially measured with three items, they were analyzed separately to maintain construct clarity: ``I can freely choose my development environment and tooling'' (56.7\% endorsement), ``My employer does not impact my IDE choice'' (34.8\% endorsement), and ``I prefer open-source IDEs I can customize to my liking'' (38.1\% endorsement). The experimenter construct relies on a single item (``I love to experiment and often change my development environment''), precluding reliability assessment. We provide behavioral validation (experimenters use 13\% more technologies, t = 5.38, p < 0.001), but single-item measures increase measurement error. Future surveys should include multiple items with Likert-scale responses to enable proper reliability assessment and construct validation following established scale development procedures.

\paragraph{Task-level Outcomes:} We did not collect task-level activity or performance measures, limiting direct linkage between autonomy and activity outcomes emphasized in pandemic/WFH research and SDT/IJARS. Future work should add task-level metrics (e.g., time-on-task, interruptions, flow) to test whether autonomy-driven satisfaction effects translate into observable productivity patterns.

\paragraph{NPS Methodology Limitations:} While NPS provides industry-standard benchmarking, the methodology has documented limitations that affect actionability and precision \citep{fisher2018nps}. NPS ignores 33.1\% of respondents (passives, scores 7--8) despite meaningful satisfaction differences, uses arbitrary cutoffs for promoter classification, and reduces rich 0--10 satisfaction data to three categories. Our supplementary Customer Value Analysis addresses some limitations by providing more granular segmentation, but researchers should be cautious about over-interpreting NPS as a precise metric.

\paragraph{Temporal Specificity:} The July 2022 timing captures a specific moment but cannot address subsequent rapid market changes.

\paragraph{Geographic Limitations:} The English-only survey excludes non-English-speaking developers, limiting global representativeness.

\paragraph{Methodological Differences:} Our single-country sample and newsletter recruitment differ from global surveys (Stack Overflow: 73,268 responses; JetBrains: 29,000 responses) with organic or multi-channel recruitment. While our confidence intervals provide statistical precision absent in industry surveys, our narrower geographic scope may limit generalizability. However, the convergent validity demonstrated through IDE usage comparisons suggests our findings capture patterns consistent with global developer populations.

\section*{Future Research Directions}

Our findings suggest several promising research directions, particularly in the context of rapid AI tool adoption:

\paragraph{Longitudinal AI Impact Studies:} Our dataset provides a baseline for tracking how AI tools have changed developer satisfaction, autonomy, and cloud adoption patterns. Researchers could replicate our methodology post-AI to measure disruption across SPACE framework dimensions, particularly testing whether the stable autonomy-satisfaction association ($\beta$ = 0.51) persists with AI tool integration.

\paragraph{Autonomy in AI Era:} Do AI tools enhance or diminish developer autonomy? Research should examine whether AI assistance increases perceived autonomy by reducing cognitive load, or decreases it by constraining solution approaches. The autonomy effect reported here provides a benchmark: if AI tools significantly disrupt this relationship, it would indicate fundamental shifts in developer experience.

\paragraph{Experience and AI Adoption:} Experimentation tendency rises with experience (32\% for 5+ years vs 23\% for juniors), suggesting nuanced AI adoption patterns. Future work should test whether experienced developers' tool knowledge makes them more discerning AI users or more resistant to AI-driven workflow changes. The 1.88-point satisfaction gap between juniors and seniors offers a baseline for whether AI tools widen or narrow experience-based satisfaction differences.

\paragraph{Technology Adoption Patterns:} The experimenter segment (29.9\%) could drive AI tool uptake, but their higher technology use does not translate into higher satisfaction. This implies AI adoption may not automatically improve developer experience. Research should examine whether AI tools alter experimenter–non-experimenter satisfaction dynamics or simply shift their tool mix.

\paragraph{AI Tool Retention Dynamics:} The established retention baseline (VS Code: 68.5\%, Eclipse: 3.9\%) provides reference points for measuring AI tool loyalty. Will AI assistants achieve higher retention than traditional IDEs? How will retention patterns shift as tools like Cursor (launched 2024) and Claude Code CLI gain adoption? These questions merit investigation as AI tools reshape the IDE landscape.

\paragraph{Organizational Strategies:} How can organizations optimize productivity while maintaining developer autonomy and satisfaction? Our autonomy penalty (0.71 points) provides quantitative evidence for satisfaction costs of tool mandates, suggesting organizations should carefully consider developer choice in AI tool implementation strategies.

\paragraph{Methodological Advances:} Future research should develop validated multi-item scales for autonomy, cognitive load, and flow to enable more precise measurement of AI impacts. Our single-item limitations highlight the need for robust measurement instruments as AI tools reshape development workflows.

Recent research suggests that coding experience influences how developers perceive AI tools, with novices often viewing AI as a "teacher" while seniors treat it as a "colleague" or no help \citep{novielli2023llmse}. These nuanced adoption patterns merit further investigation as experience levels shape AI tool acceptance and usage patterns.

Exploratory logistic regression on our current data (barrier indicators predicting cloud IDE adoption intentions) is feasible and shows modest signal (e.g., training/onboarding cost concerns OR $\approx$ 1.6; pseudo $R^{2} \approx 0.02$), motivating future work to refine barrier measures and validate richer multivariate models.

\section*{Data Availability and Reproducibility}

An anonymized dataset, complete analysis scripts, and the full survey instrument will be publicly released to ensure reproducibility and re-analysis. De-identification procedures include: removal of direct identifiers (e.g., email); hashing of internal network identifiers; coarsening of rare categories; and small-cell suppression for country-by-role cross-tabs where $n<5$. The released code reproduces all figures, tables, and robustness checks from the anonymized data.

Supplementary materials include:
\begin{itemize}[noitemsep]
    \item \textbf{Survey Instrument:} Complete questionnaire with exact item wording
    \item \textbf{Data Dictionary:} Variable descriptions and coding/collapsing rules
    \item \textbf{Analysis Scripts:} Code to reproduce all statistics and figures
    \item \textbf{Data Provenance:} Cross-checks verifying reported statistics against the released data
\end{itemize}

All materials are released under a permissive license for research use. No individual-level information that could reasonably identify respondents is included in the public release.

\section*{Conclusions}

This survey provides a comprehensive snapshot of developer tool satisfaction and autonomy immediately before the AI revolution. With high overall satisfaction (NPS = 34.7), clear experience effects (1.88-point junior-senior gap), and meaningful autonomy impacts (0.71-point satisfaction difference), these findings establish important benchmarks for the field.

The identification of distinct developer segments, including the ``experimenter'' group (29.9\%), adds nuance to our understanding of developer behavior. The relatively low cloud IDE adoption rate (4.3\%) despite industry enthusiasm highlights the practical barriers to tool adoption, with network dependency emerging as the dominant individual developer concern.

The retention analysis reveals significant underlying dynamics in IDE loyalty, with traditional tools facing existential churn risks despite moderate satisfaction. This finding, combined with the experimenter segment's high technology adoption, suggests the developer tool market was primed for disruption even before AI tools became mainstream.

Taken together, these results offer an empirical snapshot of developer attitudes and tool usage in the final period before AI-based development assistants became widely integrated into workflows. We caution against causal interpretations but note that the dataset may serve as a useful reference for studying how IDE satisfaction, autonomy, and cloud adoption evolve as tooling changes.

\section*{Acknowledgments}

We thank all developers who participated in our survey, generously sharing their time and insights. We also appreciate the helpful feedback from colleagues who reviewed early drafts of this manuscript. We are grateful to Codeanywhere for their support and collaboration. No external funding was received for this study.

\bibliography{references}

\end{document}